\documentclass[11pt]{article}

\usepackage[utf8]{inputenc}
\usepackage[T1]{fontenc}

\usepackage{amsmath,amssymb,amsthm}
\usepackage{mathtools}

\usepackage{graphicx}
\usepackage{float}

\usepackage{booktabs}
\usepackage{multirow}
\usepackage{tabularx}
\usepackage{array}

\usepackage{xcolor}
\usepackage{hyperref}
\usepackage{cite}

\usepackage[a4paper,margin=2.5cm]{geometry}

\graphicspath{{Fig/}}

\hypersetup{
    colorlinks=true,
    linkcolor=blue,
    citecolor=blue,
    urlcolor=blue
}

\theoremstyle{definition}

\theoremstyle{remark}

\title{
Scalable vertex‑guided filtrations identify structurally relevant genes in cancer networks
}

\author{
Edmara Viana$^{1,*}$,
Rodrigo Henrique Ramos$^{1,2}$,\\
Flávia Raquel Gonçalves Carneiro$^{3,4,5}$,
Cynthia de Oliveira Lage Ferreira$^{1}$
}

\date{}

\begin{document}

\maketitle


\begin{center}

{\small

$^1$ Institute of Mathematical and Computer Sciences (ICMC), University of São Paulo (USP), São Carlos, Brazil

$^2$ Federal Institute of São Paulo (IFSP), São Carlos, Brazil

$^3$ Center for Technology Development in Health (CDTS), FIOCRUZ, Rio de Janeiro, Brazil

$^4$ Brazilian National Cancer Institute (INCA), Rio de Janeiro, Brazil

$^5$ Oswaldo Cruz Institute (IOC), FIOCRUZ, Rio de Janeiro, Brazil

\vspace{0.3cm}

$^*$ Corresponding author:

\texttt{edmaraviana@usp.br}
}

\end{center}


\begin{abstract}
Topological data analysis (TDA) has established itself as a useful tool for capturing multiscale structures in complex networks, such as connected components, cycles, and cavities. Although Vietoris-Rips (VR) filtering is widely used in network analysis, it tends to be computationally expensive, especially for large networks. This work explores vertex function-based (VFB) filtering based on network measures, applying persistent homology to identify relevant topological structures in cancer-associated protein networks, and compares its effectiveness with the VR approach. The results show that VFB reproduces the second-order structures (Betti-2) identified by VR, recovering previously reported essential genes. In addition, VFB detected new driver genes, confirmed in databases such as IntOGen and NCG, and allowed analysis of third-order structures (Betti-3) that was not feasible with VR. Thus, VFB represents a scalable alternative to VR, preserving biological interpretability and complementing classical network metrics.
\end{abstract}

\vspace{0.3cm}



\section{Introduction}
Biological interaction networks, such as pathways, protein-protein interaction (PPI) networks, and regulatory networks, can be modeled as graphs, where vertices represent genes/proteins and edges represent regulatory, functional, or physical relationships. In the context of cancer, these networks help characterize how gene alterations propagate through cellular processes and prioritize genes with potential prognostic and therapeutic relevance \cite{barcelos, pop, WU, Koh, RamosA}. In particular, cancer can be understood as a disease in which genomic mutations accumulate and deregulate essential processes, motivating the distinction between \emph{driver} mutations (that have functional impact and contribute to tumor progression) and \emph{passenger} mutations (which do not confer a selective advantage during cancer development and progression). In this scenario, it makes sense to search for genes whose mutations cause relevant structural changes in the network (e.g., in topological cavities) as these genes are more likely to correspond to driver genes or have a meaningful association with cancer \cite{ataqueramos, TDAgenomics}.

The analysis of these networks is often based on classical graph metrics (degree, centralities, modularity), which are useful for identifying hubs and modules. However, such metrics can be sensitive to noise and incompleteness, and they capture multiscale and high-order patterns in a limited way. This hinders robust comparisons between networks and the detection of relevant global structures in cancer subtypes \cite{naderi, TDAgenomics}.

To overcome these limitations, Topological Data Analysis (TDA) provides a complementary alternative: through Persistent Homology (PH), it track the appearance and disappearance of connected components, cycles, and cavities throughout a filtration, producing barcodes and persistence diagrams that summarize the system's organization at multiple scales \cite{carltdacancer, bukkri, ataqueramos, otter}. In oncological applications, this approach has revealed patterns and potential biomarkers not easily captured by traditional descriptors \cite{ataqueramos, loughrey2021topology}.

Despite its potential, computing persistent homology in networks faces computational bottlenecks. The most common strategy in the literature is Vietoris-Rips (VR) filtering, whose number of simplices can grow combinatorially, becoming prohibitive in larger networks \cite{bukkri, ataqueramos, edelsbrunner2008computational, zomorodian}. An alternative is to use filtrations guided by network properties (e.g., attributes of vertices and edges, centrality measures), which can reduce cost while emphasizing interpretable structures  \cite{tdaemredesAktas}.

In this work, we investigate vertex-function-based (VFB) filtrations, constructed from vertex attributes (centralities), and compare their ability to (i) reproduce relevant topological structures and (ii) highlight genes associated with persistent cycles and cavities in cancer-related protein networks. Our goal is to evaluate the practical usefulness of these filtrations as a scalable alternative to Vietoris-Rips, maintaining biological interpretability and complementing classical network metrics.

To this end, we analyze the topological relevance of genes present in cancer consensus networks (CCNs) associated with three superpathways: Programmed Cell Death (PCD), DNA Repair (DR), and Chromatin Organization (CO). We adopted a methodology similar to \cite{ataqueramos}, replacing VR with VFB filtration.

Our results indicate that VFB filtrations reproduce VR findings for second-order topological structures ($\beta_2$). Furthermore, VFB reveals additional topologically relevant genes with significant biological implications in cancer and enables analysis of the topological impact of genes in three-dimensional structures ($\beta_3$), which was infeasible with VR. In summary, VFB replicates VR results for $\beta_2$, enables systematic analysis of $\beta_3$, and reveals additional driver genes.

\section{Mathematical background}\label{sec:background}
In this section, we gather only the necessary definitions of complex networks and TDA to support the methodology based on filtering guided by vertex attributes and persistent homology \cite{otter, TDAgenomics, edelsbrunner2008computational}.

\subsection{Networks as graphs}
We model a network as a graph $G=(V,E)$, where $V$ is the set of vertices (genes/proteins) and $E$ is the set of edges (interactions). In this work, the edges are undirected and unweighted.

The adjacency matrix $A=(A_{ij})$ encodes connectivity, and the degree of a vertex $i$ is $k_i=\sum_j A_{ij}$ (in the unweighted case). A path is a sequence of vertices connected by edges, and the distance $d_{ij}$ is the length of the shortest path between $i$ and $j$. A connected component is a subgraph in which every pair of vertices is connected by some path, in the analyses in this work, we focus on the largest connected component to reduce artificial fragmentation effects.

\subsection{Centralities measures}
To induce our analysis, we use vertex attributes that summarize different notions of structural importance. In general terms: (i) degree quantifies local connectivity; (ii) betweenness centrality measures intermediation in shortest paths; and (iii) eigenvector centrality assign greater importance to vertices that are connected to equally important neighbors.

\subsection{Simplicial complexes and clique complexes}
TDA studies topological invariants of combinatorial objects derived from data. A \emph{simplicial complex} is a collection of simplices (vertices, edges, triangles, tetrahedra, etc.) that is closed under taking faces. In the context of networks in this work, we use the \emph{clique complex} (or \emph{flag complex}): given a graph, a $(k-1)$-simplex is added for each clique with $k$ vertices. This construction allows us to represent higher-order interactions implicitly present in the connectivity of the graph.

\subsection{Filtrations and persistent homology}
A \emph{filtration} is an increasing sequence of simplicial complexes
$\emptyset = K_0 \subseteq K_1 \subseteq \cdots \subseteq K_n = K$.
Persistent homology (PH) tracks, throughout this sequence, the appearance and disappearance of connected components, cycles, and cavities. More precisely, let $\{K_\alpha\}_{\alpha \in \mathbb{N}}$ be a filtration of simplicial complexes. For each $k \geq 0$, consider the chain group $C_k(K_\alpha)$, defined as the vector space over a field $\mathbb{F}$ generated by the $k$-simplices of $K_\alpha$. The boundary operators
$
 \partial_k : C_k(K_\alpha) \to C_{k-1}(K_\alpha)
$ 
 are linear maps defined on a $k$-simplex $[v_0,\dots,v_k]$ by
 \begin{equation*}
 \partial_k [v_0,\dots,v_k] = \sum_{i=0}^k (-1)^i [v_0,\dots,\hat{v}_i,\dots,v_k],
\end{equation*}
 and they satisfy $\partial_k \circ \partial_{k+1} = 0$.
 The $k$-th homology group is then defined as
 \begin{equation*}
      H_k(K_\alpha) = \frac{ Z_k(K_\alpha)}{B_k(K_\alpha)},
 \end{equation*}
 where $Z_k = \ker \partial_k$ (cycles) and $B_k = \operatorname{im} \partial_{k+1}$ (boundaries). Intuitively, homology captures topological features that persist beyond local connectivity, providing a global description of the network structure that is robust to small perturbations. In this context, PH tracks, along the filtration, the birth and death of classes in $H_0, H_1, H_2$, and so on.  These features are summarized  by the Betti numbers $\beta_k(\alpha) = \dim H_k(K_\alpha)$\cite{ TDAgenomics, edelsbrunner2008computational}.

 Persistence barcodes and persistence diagrams are visual summaries of persistent homology. Each homology class that appears at a parameter value $b$ and disappears at $d$ is represented either as an interval $(b,d)$ in a barcode or as a point $(b,d)$ in a persistence diagram. These representations compactly encode the birth-death structure of topological features along the filtration.

\subsection{Filtrations on networks}
Let $G$ be a network endowed with a distance function $d$ on the vertex set (e.g., shortest-path distance). For each threshold $\varepsilon \ge 0$, define the threshold graph $G_\varepsilon$ with the same vertex set of $G$ and edges $(u,v)$ whenever $d(u,v) \le \varepsilon$.
The \emph{Vietoris-Rips complex} $\mathrm{VR}_\varepsilon$ is then defined as the clique complex of $G_\varepsilon$, i.e., a $k$-simplex is included whenever its vertices form a $(k+1)$-clique in $G_\varepsilon$.
As $\varepsilon$ increases, the complexes $\{\mathrm{VR}_\varepsilon\}_{\varepsilon \ge 0}$ form a nested filtration capturing higher-order connectivity patterns in the network
\cite{zomorodian}. Although natural from a metric viewpoint, this construction may lead to combinatorial growth in the number of simplices as the threshold increases.

In contrast, a \emph{vertex-function-based (VFB) filtration} is induced by a function $g: V \to \mathbb{R}$, which is extended to simplices by the monotone rule $f(\sigma) = \max_{v \in \sigma} g(v)$ (sublevel filtration). For each threshold $\alpha$, one considers the subgraph induced by vertices with $g(v) \le \alpha$ and its associated clique complex. This approach leverages intrinsic structural properties of the network (e.g., centrality measures) and typically reduces computational cost by restricting the construction to cliques already present in the graph \cite{tdaemredesAktas}.

\begin{figure}[!t]
\centering
\includegraphics[width=\linewidth]{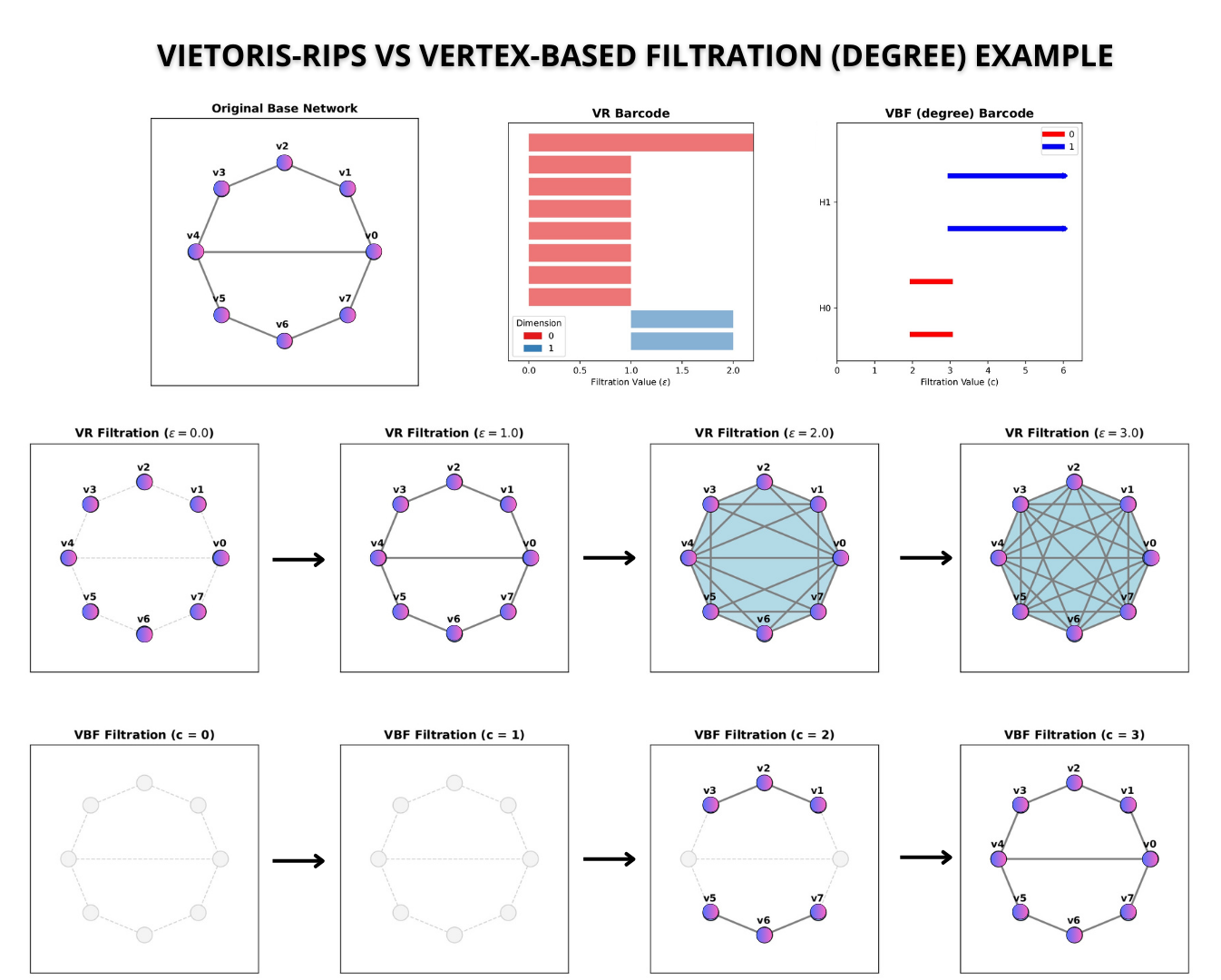}
\caption{Schematic comparison between Vietoris-Rips (VR) filtering and vertex function-based (VFB, degree) filtering on a fictitious network. The top row shows the original network and the resulting persistence barcodes for VR (left) and VFB (right). The bottom panels illustrate successive VR filtrations with increasing distance limits and VFB filtrations with increasing degree limits.}
\label{VFB}
\end{figure}

To provide an intuitive illustration of these two constructions, Figure \ref{VFB} compares the Vietoris-Rips filtration and a vertex-function-based (degree) filtration on a small example network, showing the evolution of the underlying graphs and the corresponding persistence barcodes.

\subsection{Vertex removal}
To assess the sensitivity of topological descriptors to structural perturbations, we generate perturbed versions of the network by removing vertices individually. In each iteration, a single vertex (gene/protein) and its associated edges are removed, while all other vertices and edges remain unchanged. Consequently, each perturbed network differs from the original by the deletion of a single gene.

Persistent homology is recomputed for every such single-vertex perturbation, and the resulting invariants are compared with those of the original network. As our primary interest lies in higher-order organization, we focus on the effects on the higher-dimensional Betti numbers, particularly $\beta_2$ and $\beta_3$, thereby quantifying the individual contribution of each gene to two- and three-dimensional topological structures.
 This procedure follows the same approach adopted in the work of \cite{ataqueramos}. 


\section{Methodology}\label{sec:methods}
We used networks previously constructed and analyzed in \cite{ataqueramos}, obtained through the integration of somatic mutation data with superpathway networks.

The Cancer Genome Atlas (TCGA)\footnote{TCGA data are fully deidentified and anonymised. Consequently, ethical approval was not required for this study.} is a resource that provides access to comprehensive datasets containing detailed information on various types of cancer mutations. The Mutation Annotation Format (MAF) is a tab-delimited file widely used in the field and serves to link patient samples, genes, and mutations.

Superpathways consist of sets of genes that cooperate to perform specific biological functions and correspond to subnetworks of the protein-protein interaction network (PPIN) curated in the Reactome Knowledgebase\footnote{Reactome: \url{https://reactome.org}} \cite{WU}. In practice, Reactome represents pathways as gene lists, which enables the extraction of induced subgraphs from the PPIN to construct superpathway networks (SPNs). The mutation lists provided in MAF files are then projected onto these SPNs and their interactions, yielding comparable cancer consensus networks (CCNs) across cancer types \cite{ataqueramos}.

As in \cite{ataqueramos}, we integrated mutation data from six cancer types and focused on three biological functions, thereby obtaining three CCNs of interest: DNA  Repair (DR-CCN), Chromatin Organization (CO-CCN), and Programmed Cell Death (PCD-CCN). In all analyses, we consider only the largest connected component in order to avoid topological artifacts induced by network fragmentation.

\begin{figure}[!t]
\centering
\includegraphics[width=\linewidth]{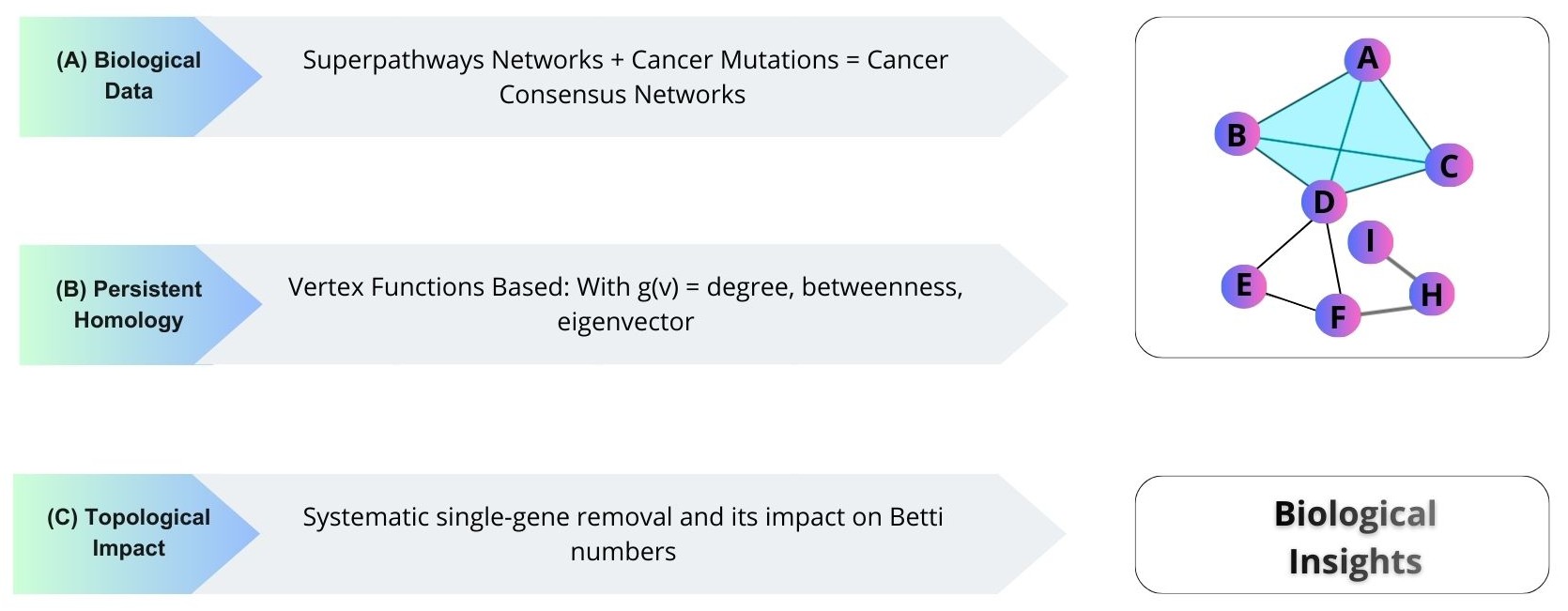}
\caption{Overview of the topological analysis pipeline for CCNs. (A) Integration of PPIN, superpathways, and MAF data from six cancer types to build CCNs. (B) Vertex-function-based (VFB) filtration using centrality measures to compute persistent homology and Betti numbers ($\beta_0$, $\beta_1$, $\beta_2$, $\beta_3$). (C) Single-gene knockout analysis to assess topological impact (e.g., removal of genes A, E, H, I). Only the largest connected component is analyzed to avoid fragmentation artifacts.}
\label{pipeline}
\end{figure}

Figure \ref{pipeline} illustrates the overall methodology. (A) This step summarizes a more detailed procedure described in \cite{ataqueramos}, in which a PPIN, superpathways, and mutation data from six cancer types (MAFs) are integrated to construct the CCNs. (B) Vertex centrality measures are used to define VFB filtrations, enabling the detection of higher-dimensional topological structures. The combination of steps (A) and (B) results in the identification of the Betti numbers present in the CCNs, as illustrated in the figure. (C) We then perform systematic single-gene removal to quantify the impact of each gene on the network’s topological structures and to infer the biological relevance of those that significantly affect them. In the illustrative network shown in the figure, removing gene I does not produce measurable changes in the Betti numbers. Removing gene H creates additional connected components, thereby affecting $\beta_0$. Removing gene E preserves global connectivity but disrupts a cycle, impacting $\beta_1$. Finally, in the case in which the tetrahedron determined by A, B, C, and D does not constitute a 3-simplex, the removal of gene A preserves network connectivity while eliminating one two-dimensional void ($\beta_2$).

With this framework in place, we now describe the concrete computational steps implemented using \texttt{GUDHI}, focusing on $\beta_2$ and $\beta_3$.

The resulting CCNs were subjected to different VFB filtration strategies driven by vertex attributes. Specifically, we considered the following centrality measures as filtering functions: degree, betweenness and eigenvector centrality.

For each CCN and each chosen vertex attribute, we constructed the corresponding vertex-function-based filtration of simplicial complexes and computed the associated persistent homology invariants. The computations were performed with coefficients in $\mathbb{Z}_2$ using the \texttt{GUDHI} library \cite{Gudhi}, producing persistence diagrams, barcodes, and numerical summaries. Our analysis focuses on higher-dimensional Betti numbers, in particular $\beta_2$ and $\beta_3$, when present, in order to characterize two- and three-dimensional topological structures in cancer-related protein networks.

We also evaluated perturbed versions of the networks obtained by single-vertex removal in order to assess the robustness of the identified topological patterns. This procedure further allows us to quantify the individual impact of each gene on two- and three-dimensional topological structures (as measured by  $\beta_2$ and  $\beta_3$), thereby revealing a notion of topological relevance that can be related to its biological importance, as discussed below.

\section{Results}\label{sec3}

In this section, we evaluate the effectiveness of VFB filtering in identifying topologically significant genes in CCNs. We compare our results with the traditional Vietoris-Rips (VR) approach used by \cite{ataqueramos}, focusing on computational efficiency, recovery of known biological factors, and discovery of higher-order structures.

\subsection{Consistency and validation of VFB filtering}

First, we assess whether VFB filtering can recover the topological features identified by the VR approach, which is more computationally demanding. Our results show high consistency: using degree, betweenness, and eigenvector centrality measures as vertex function $g(v)$, we successfully recovered almost all genes previously reported as having a significant impact on $\beta_2$ in the previous work of \cite{ataqueramos}.

Specifically, for DR-CCN (233 nodes), CO-CCN (162 nodes), and PCD-CCN (170 nodes), degree and betweenness based filters identified the full set of essential genes described in \cite{ataqueramos}. Table \ref{quadro} summarizes the essential genes impacting $\beta_2$ in each CCN, as recovered by degree and betweenness VFB filtrations.

\begin{table}[!t]
\centering
\caption{Genes that impact $\beta_2$ in CCNs, identified by VFB filtering. Known driver genes are highlighted in bold.}
\label{quadro}
\noindent\begin{tabular}{|c|p{0.75\columnwidth}|}
\hline
CO-CCN & ACTL6A, BRMS1, \textbf{RELA, SMARCE1,} WDR77 \\\hline
DR-CCN &  \textbf{ABL1}, ACTL6A, \textbf{ATM, ATR, EP300, FANCD2}, HERC2, KAT5, PCNA, POLN, RAD51, \textbf{XPA},  XRCC6 \\\hline
PCD-CCN & \textbf{AKT1}, APAF1, BAD, BIRC2, CASP1, CASP3, CASP6, CASP8, \textbf{CTNNB1}, HSP90AA1, MAPT, PTK2, \textbf{RIPK1}, ROCK1, \textbf{STAT3}, STUB1, TNFSF10, \textbf{TP53} \\\hline
\end{tabular}
\end{table}

Interestingly, the eigenvector centrality measure failed to capture only the \textit{PCNA} gene in the DR-CCN network, suggesting that, while most centralities are robust, the choice of vertex function can subtly adjust sensitivity to specific biological hubs.

\subsection{Discovery of additional topologically relevant genes}

In addition to validation, the VFB approach provides a more nuanced view of the higher-order organization of the networks. Using the intrinsic structure of the graph, we identified several genes that affected $\beta_2$ but were not detected by the VR method. 

\begin{figure}[!t]
\centering
\includegraphics[width=\linewidth]{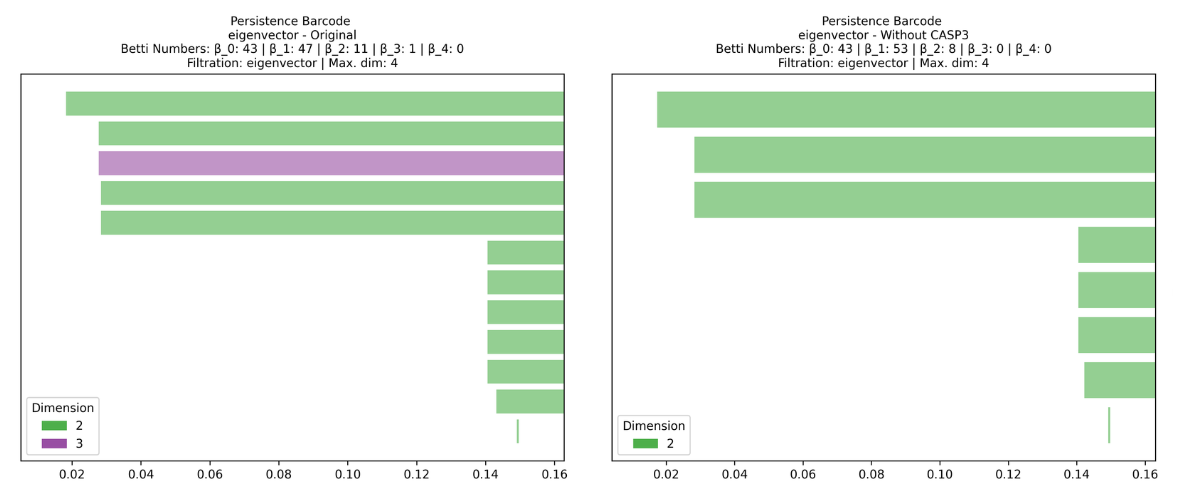}
\caption{Persistence barcodes for the original PCD-CCN (left) and the same network with removal of gene \textit{CASP3} (right). The disappearance of bars of dimensions 2 and 3 indicates the disruption of two-dimensional and three-dimensional topological cavities, reflecting the topological relevance of the removed gene.}
\label{barcode}
\end{figure}

To illustrate how the removal of a single gene affects higher‑order topological structures, we compare persistence barcodes computed from the original network and from the network in which a single vertex (gene) is removed. Figure \ref{barcode} shows the barcode for the original PCD-CCN and for the perturbed network in which gene \textit{CASP3} is deleted. The left panel corresponds to the unperturbed network, while the right panel shows the barcode after node removal. The  disappearance of  bars in dimension 2 and 3 indicates disruption of two‑ and three‑dimensional topological cavities, which we interpret as evidence of topological relevance of gene \textit{CASP3}. 

To assess the clinical and biological importance of these new candidates, we cross-referenced them with the IntOGen\footnote{IntOGen: \url{https://www.intogen.org/search}} and Network of Cancer Genes (NCG)\footnote{{NCG: \url{http://network-cancer-genes.org/index.php}}} databases. This analysis confirms that these genes are classified as \textit{drivers}, reinforcing that VFB screening is not only more efficient but also more sensitive in prioritizing genes involved in tumor progression. Tables \ref{tab1}, \ref{tab2} and \ref{tab3} list the driver genes detected by VFB, but not by VR.

\begin{table}[!t]
\centering
\caption{Driver genes impacting $\beta_2$ in the CO-CCN network, identified by VFB filtrations and not detected by the VR approach.}
\label{tab1}
\small
\begin{tabular}{lccc}
\toprule
Genes & Degree & Betweenness & Eigenvector   \\
\midrule \\
ARID1A & X & X & X \\
CCND1  & X & X & X \\
CDK4   & X &  & X \\
CREBBP &  &  & X \\
EDD   &  & X & \\
EP300  &  &  & X \\
NCOA1  & X & X & X \\
REST   & X & X & \\
SETDB1 &  &  X &  \\
SMARCA2 & X & X & X \\
SMARCA4 & X & X & X \\
SMARCD1 & & X & \\
TRRAP   & X & X & X \\
\toprule
\end{tabular}
\end{table}

\begin{table}[!t]
\centering
\caption{Driver genes impacting $\beta_2$ in the DR-CCN network, identified by VFB filtrations and not detected by VR approach.}
\label{tab2}
\small
\begin{tabular}{lccc}
\toprule
Genes & Degree & Betweenness & Eigenvector   \\
\midrule \\
BLM  &  & X & \\
BRCA1& X &  & X \\
CHEK2& X &  & \\
ERCC4 & X &  &  \\
MSH2 & X  &  X & X \\
NTHL1 & & X & \\
TP53 & &  & X \\
WRN & &  & X \\
\toprule
\end{tabular}
\end{table}

\begin{table}[!t]
\centering
\caption{Driver genes impacting $\beta_2$ in the PCD-CCN network, identified by VFB filtrations and not detected by VR; betweenness and eigenvector centralities did not reveal additional drivers in this case. }
\label{tab3}
\small
\begin{tabular}{lccc}
\toprule
Genes & Degree & Betweenness & Eigenvector   \\
\midrule \\
AKT3  & X &  &  \\
FAS    & X &  & \\

\toprule
\end{tabular}
\end{table}

One of the main limitations of the classic VR approach is its combinatorial explosion, which often makes the computation of three-dimensional features ($\beta_3$) infeasible for networks of this size. Due to the reduced computational cost of VFB, we were able to perform a complete systematic single-gene removal analysis for $\beta_3$. We found genes whose removal disrupts three-dimensional topological cavities, suggesting that they act as ``structural anchors'' for complex protein-protein interaction clusters.

Tables \ref{tab4} and \ref{tab5} summarizes these findings. Detecting these $\beta_3$ structures adds a new layer of evidence about the multiscale organization of cancer pathways.

\begin{table}[!t]
\centering
\caption{Genes whose removal impacts three-dimensional topological structures ($\beta_3$) in the DR-CCN, identified using VFB filtrations. Known driver genes are highlighted in bold.}
\label{tab4}
\small
\begin{tabular}{lccccccc}
\toprule
Gene & Degree & Betweenness & Eigenvector  \\
\midrule \\
\textbf{BARD1} &  & x &  &  \\
DDB1 & x & x & x &   \\
\textbf{ERCC4} & x &  & x &   \\
PARP1 & x &  & x &   \\ 
PCNA  & x &  & x &   \\
\textbf{TP53 } &  & x &  &   \\
\textbf{WRN}  & x &  & x &   \\
\toprule
\end{tabular}
\end{table}

\begin{table}[!t]
\centering
\caption{Genes whose removal impacts three-dimensional topological structures ($\beta_3$) in the PCD-CCN, identified using VFB filtrations. Known driver genes are highlighted in bold.}
\label{tab5}
\small
\begin{tabular}{lccccccc}
\toprule
Gene & Degree & Betweenness & Eigenvector  \\
\midrule \\
APAF1 & x & x & x &  \\
BAD & x & x & x &   \\
\textbf{BAX }& x & x & x &   \\
BCL2L1 & x & x & x &   \\ 
\textbf{CASP3} & x & x & x &   \\
CASP9  & x & x & x &   \\
MAPK3  & x & x & x &   \\
MAPK8  & x &  x& x &   \\
\toprule
\end{tabular}
\end{table}

Among the genes identified, \textit{BARD1}, \textit{ERCC4}, \textit{TP53} and \textit{WRN} appear in the DR-CCN network, and \textit{BAX} and \textit{CASP3} in the PCD-CCN network, all classified as driver genes. The CO-CCN network did not possess any three-dimensional structures.

These results suggest that vertex attribute guided filtering preserves previously described biological signals while offering flexibility to emphasize different notions of node importance.

In the following section, we will discuss the role and biological relevance in cancer of the genes that impacted the third Betti number ($\beta_3$) according to our methodology.


\subsection{Biological significance of $\beta_3$-influencing genes in cancer}

Using the proposed methodology, we identified key genes associated with high-dimensional network structures, notably those associated with third Betti number ($\beta_3$) features.
The network structures captured by $\beta_3$ are associated with genes that play key roles in two major cancer‑related pathways. In the following paragraphs, we highlight how these genes contribute to programmed cell death and DNA repair, respectively.

\subsubsection{Relevance of $\beta_3$-influencing genes in PCD}
Apoptosis is a tightly regulated process that maintains tissue homeostasis by eliminating damaged or unnecessary cells without triggering inflammation. Our method identified, within the PCD network, key genes, such as \textit{BAD, BAX, BCL2L1, CASP3}, and \textit{CASP9}. Although \textit{MAPK3} and \textit{MAPK8} are implicated in apoptosis in specific cellular contexts, their roles as pro-survival factors in tumorigenesis are well established.

\textit{BAX} and \textit{BAD} are pro-apoptotic members of the BCL-2 family and cytochrome \textit{c} release. \textit{BAX} low expression correlates with poor prognosis in several tumors \cite{Gutta}, while \textit{BAD} expression correlates with positive clinical outcomes. 
However, \textit{BAD} is a multifunctional protein that has also been linked to cell growth in cancer cells. It can be either pro-apoptotic or pro-survival depending on the phosphorylation site \cite{Mann}.
In contrast, \textit{BCL2L1} is a critical anti-apoptotic protein. Its overexpression has been reported in various types of cancer, 
inducing cancer progression and metastasis \cite{Silva}.

\textit{CASP9} and \textit{CASP3} encode the caspase-9 and -3, respectively. These are fundamental proteins that function in the apoptosis mechanism. Caspase-9 is the initiator caspase of the mitochondrial pathway, while caspase-3 is an effector caspase that is activated by proteolytic processing and, in turn, cleaves vital cellular proteins, leading to cell death \cite{Kiraz}. Targeting caspases with inducer molecules opens a new perspective of cancer treatment. 

\textit{APAF1} is a central regulator of the intrinsic apoptotic pathway, responsible for apoptosome formation and activation of caspase-9 following mitochondrial cytochrome \textit{c} release. Its reduced expression or functional impairment in cancer cells contributes to apoptosis evasion and therapy resistance, supporting tumor progression \cite{Soengas}.

Finally, \textit{MAPK3}, also known as \textit{ERK1}, and \textit{MAPK8}, also known as \textit{JNK1}, have dual roles in apoptosis. Although these genes are widely associated with anti-apoptotic functions by regulating cell proliferation, differentiation, and survival, 
they can be pro-apoptotic in some contexts \cite{Abizadeh, Wu2019}.

\subsubsection{Relevance of $\beta_3$-influencing genes in DNA repair}
The DNA repair pathway, a fundamental cellular mechanism responsible for maintaining genomic stability, also exhibited key genes whose removal impacted $\beta_3$ structures.

\textit{WRN} is  essential for genome stability, DNA repair, and replication, and is highly expressed in tumor cells. Its silencing selectively induces mitotic catastrophe and death in cancer cells while sparing normal cells, making \textit{WRN} a promising therapeutic target. 
Inhibition of \textit{WRN} also enhances the efficacy of genotoxic chemotherapy, underscoring its importance in cancer treatment strategies \cite{futami}. 

\textit{ERCC4}, a key gene in the nucleotide excision repair pathway, has been significantly associated with cancer susceptibility in a large meta-analysis of 160 studies. In particular, the \textit{ERCC4} rs744154 polymorphism showed strong evidence of association with increased bladder cancer risk in Asian populations, with functional annotations suggesting a potential regulatory role. These findings highlight the importance of \textit{ERCC4} in modulating genetic predisposition to cancer \cite{zuo}. 

\textit{PARP1} is a critical enzyme involved in several signalling pathways related to DNA repair and is upregulated in many tumours, especially breast, uterine, lung, ovarian, and skin cancers, as well as non-Hodgkin’s lymphoma \cite{conceicao}. 

\textit{PCNA} has a fundamental role in DNA replication and repair, serving as a processivity factor for DNA polymerase to maintain genomic integrity and prevent the propagation of DNA errors. It is frequently overexpressed in highly proliferating tumor cells and has a role in maintaining cancer stemness \cite{Wang}.

Beyond core DNA repair enzymes, our analysis also highlighted genes involved in genome surveillance and damage response regulation.
\textit{BARD1} (\textit{BRCA1} Associated RING Domain 1) encodes a protein that forms a heterodimer with BRCA1. This interaction is critical for the complex’s E3 ubiquitin ligase activity, which is important for \textit{BRCA1} tumor suppressor function. Despite its well-characterized role in genome maintenance, \textit{BARD1} plays a multifaceted role in cancer biology, acting as either a tumor suppressor or an oncogene, depending on the isoforms expressed, cellular context, and microenvironment. Interestingly, \textit{BARD1} loss-of-function variants were associated with early-onset familial breast cancer, indicating that enhanced breast cancer screening strategies should be considered for women harboring pathogenic \textit{BARD1} variants \cite{weber}.

\textit{DDB1} plays an essential role in nucleotide excision repair (NER) following ultraviolet damage. Upregulation of DDB1 was observed in several cancers and correlated with poor prognosis, such as in ovarian, colorectal, and pancreatic cancer \cite{Jo}.

\textit{TP53} is the most frequently mutated gene in human cancers. Mutant p53 not only loses tumor-suppressive activity but also acquires gain-of-function properties that drive tumor progression, genomic instability, altered ferroptosis, microenvironment remodeling, and stemness. Its presence is associated with poor prognosis, influences responses to chemotherapy and radiotherapy, and represents an important target for novel anticancer therapies \cite{Chen}.

Overall, the biological functions of genes impacting $\beta_3$ structures suggest that high-dimensional topological features encode non-trivial information about cancer-related mechanisms, motivating their investigation.
\section{Conclusion}
We investigated vertex-function-based (VFB) filtrations as a scalable alternative to Vietoris-Rips (VR) filtrations for persistent homology in cancer-related protein interaction networks. Overall, VFB recovered the key genes impacting $\beta_2$ structures reported with VR, substantially reducing computational cost and enabling systematic perturbation experiments that are impractical with VR.

In addition to reproducing known results, VFB highlighted additional genes associated with persistent higher-order structures and supported their biological relevance through cross-references to cancer driver databases. Furthermore, the feasibility of computing $\beta_3$ under VFB allowed us to identify genes that act as structural anchors for three-dimensional cavities in specific superpathway consensus networks, providing complementary evidence of multiscale organization.

Taken together, our results indicate that vertex-guided filtering preserves important topological signatures while improving efficiency and interpretability, making it a promising tool for prioritizing candidate driver genes and extending topological analyses to higher-dimensional structures in cancer network studies.

Future work includes evaluating additional vertex functions, assessing robustness across different cancer types and interaction protein networks, and integrating functional validation to connect persistent cavities to clinically relevant phenotypes.





\section*{Funding}
This study was financed in part by the Coordination of
Improvement of Higher Education Personnel (CAPES) - Brazil.


\section*{Data availability}

All code and files are available on GitHub: \url{https://github.com/edmaraviana-git/Paper---2026.git}


\section*{Author contributions statement}
E.V and C.O.L.F designed and conceptualized the study and the experiments. E.V. conducted the experiments.
E.V. and F.R.G.C. analyzed the results. 
E.V., R.H.R., F.R.G.C. and C.O.L.F. wrote and revised the manuscript. C.O.L.F coordinated the study.




\end{document}